\newcommand{\cR}{\mbox{${\cal R}$}}

\documentclass{article}

\newtheorem{theorem}{Theorem}
\newtheorem{corollary}{Corollary}
\newtheorem{proposition}{Proposition}
\newtheorem{lemma}{Lemma}
\newtheorem{claim}{Claim}

\newtheorem{definition}{Definition}
\newtheorem{example}{Example}

\begin{document}
\newcommand{\blackslug}{\mbox{\hskip 1pt \vrule width 4pt height 8pt 
depth 1.5pt \hskip 1pt}}
\newcommand{\qed}{\quad\blackslug\lower 8.5pt\null\par\noindent}
\newenvironment{proof}{\par\noindent{\bf Proof:}}{\qed}

\title{Combinatorial Auctions with Decreasing Marginal Utilities\thanks{A preliminary version of this paper has been presented at EC'01}
\thanks{Supported by grants from the Israeli Ministry of Science and the Israeli Academy of Sciences}}

\author{Benny Lehmann\\Digital Fuel\\E-mail: benny@digifuel.com
\and Daniel Lehmann\\School of Computer Science and Engineering, The Hebrew University\\Jerusalem, Israel\\E-mail: lehmann@cs.huji.ac.il
\and Noam Nisan\\School of Computer Science and Engineering, The Hebrew University\\Jerusalem, Israel\\E-mail: noam@cs.huji.ac.il
}

\date{September 12th, 2002}
\maketitle

\begin{abstract}
In most of microeconomic theory, consumers are assumed to exhibit decreasing 
marginal utilities. 
This paper considers combinatorial auctions among such {\em submodular} buyers.
The valuations of such buyers are placed 
within a hierarchy of valuations that exhibit 
no complementarities, a hierarchy that includes also 
OR and XOR combinations of singleton valuations, 
and valuations satisfying the gross substitutes property.
Those last valuations are shown to form a zero-measure subset of the submodular
valuations that have positive measure.
While we show that the allocation problem among submodular valuations is NP-hard,
we present an efficient greedy $2$-approximation algorithm for this case
and generalize it to the case of limited complementarities.
No such approximation algorithm exists in a setting allowing 
for arbitrary complementarities.  
Some results about strategic aspects of combinatorial auctions among players 
with decreasing marginal utilities are also presented.
\end{abstract}

\section{Introduction and Overview}
\subsection{Background}
\label{sec:back}
Recent years have seen a surge of interest in combinatorial 
(also called combinational) auctions,
in which a number of non-identical items are sold concurrently 
and bidders express preferences about 
combinations of items and not just about single items.  
Thus, for example, a bidder may offer \$40 for the combination 
of a Tournedos Rossini and a bottle of Chateau Lafitte, 
but offer only \$10 for each of those items alone.  
Similarly, a bidder may make an offer of \$10 for a blue and for a red shirt, 
but not be willing to pay more than \$10 even if he gets both shirts.  
In general, a combinatorial auction allows bidders 
to express complementarities -- where 
the value of a combination of packages of items is worth 
more than the sum of the values of the separate packages --
and substitutabilities -- where the value of a combination of packages 
is less than the sum of the values of the separate packages.  
Such combinatorial auctions have been suggested for a host of 
auction situations such as those for spectrum licenses, pollution permits, 
landing slots, computational resources, and others.  
See de Vries and Vohra(forthcoming) for a survey.

Implementation of combinatorial auctions faces several challenges 
including the representational question of succinctly specifying the values 
of the different packages, the algorithmic challenge of efficiently solving 
the resulting, NP-hard, allocation problem, and the game-theoretic
questions of bidders' strategies.  
These questions have been recently approached by a host of researchers
both in the general case and in several interesting special 
cases: MacKie-Mason and Varian(1994), Varian(1995), 
Rothkopf and al.(1998),
Sandholm(1999), Sandholm and Suri(1999), Fujishima and al.(1999),
Tennenholtz(2002), Lehmann and al.(forthcoming), Gonen and Lehmann(2000),
and Nisan(2000).

Somewhat surprisingly, the special case that is most often studied in economics 
has received very little attention from the computational point of view.
In most of microeconomic theory, consumers are assumed to exhibit diminishing 
marginal utilities.  
In particular, such consumers exhibit {\em no complementarities}.  
While it is not clear to what extent this assumption may be justified in practice, 
it is widely used. At least, it is mathematically convenient, but we think it is
often justified in practice.
Many papers dealing with computational issues in combinatorial auctions 
focused on the dual case of no substitutes, 
i.e., much computational research assumes that a buyer places bids for packages 
of items and is not interested in a sub-package.
In contrast, economists who dealt with auctions did mostly consider auctions 
in which players expressed no complementarities.  
For example, for multi-unit auctions, 
Vickrey's seminal paper Vickrey(1961) assumes
{\em downward sloping} valuations for buyers.
Recent papers dealing with combinatorial auctions
such as Kelso and Crawford(1982), Gul and Stacchetti(1999), 
Gul and Stacchetti(2000), Ausubel(2000) and Milgrom(2000)
usually assume the {\em gross substitutes} property.  
Each of these notions implies lack of complementarities.


Perhaps this lack of computational attention is due to a representational 
difficulty:
it is usually assumed that buyers express their preferences by non-exclusive bids
on sets of items. 
Representing buyers with no complementarities seems to require the use
of exclusive bids for sets of items (XOR-bids), 
and the mechanics of such exclusive bids have never been studied in full.
Bids on packages of items inherently express a very strong form 
of complementarity: sub-packages are not valued at all, and therefore such bids
seem to provide a very poor starting point to describe types that exhibit no
complementarities. 
The prevalent method for dealing with substitutabilities is the 
{\em phantom good} approach of Fujishima and al.(1999) 
and Nisan(2000), 
that simulates exclusive bids by non-exclusive bids on an enlarged set of goods.
This is certainly not a promising approach for studying 
the specifics of exclusive bids or types exhibiting no complementarities.  


In this paper, we study the three notions found in the literature 
for lack of complementarities. 
From the most general to the most specific these notions are, informally:
\begin{itemize}
\item {\bf No Complementarities}: 
The value of a combination of bundles is no more than
the sum of the bundle values.
\item {\bf Decreasing Marginal Utilities}: 
The marginal value of an item decreases as the
set of items already acquired increases.
\item {\bf Gross Substitutes}: 
The demand for an item does not decrease when the prices of other items increase.
\end{itemize}
This paper's main message is that the case of
{\em decreasing marginal utilities}, 
which is equivalent to submodularity of the valuation functions, should be the 
focus of interest.

We prove three types of results:

\begin{itemize}
\item {\bf Representation and Hierarchy:}
We study the question of how to represent valuations (i.e., bids) 
that exhibit no complementarities.
Our representations are constructed out of atomic valuations 
that offer a price for a single item, ({\em singleton valuations}),
using the operations of OR and XOR.  
Using restricted classes of such bidding languages 
we are able to obtain
a structural hierarchy of families of complement-free valuations.
Those {\em syntactic} classes are compared to three {\em semantic} classes.
This hierarchy contains five strictly nested classes of valuations:
(a) complement-free valuations; (b) XOR-of-OR-of-singleton valuations; (c)
submodular valuations (d) valuations that satisfy the
gross substitutes property; and (e) OR-of-XOR-of-singleton valuations.  
We prove 
the containment of each class within the previous one and
exhibit examples showing that the containment is strict.  
We also prove theorems regarding closure properties
of these classes under the operations of OR and XOR.  
We finally show that in terms of measure the significant
gap is between the class of submodular valuations and the class of valuations 
that satisfy the gross substitutes condition.

\item {\bf Allocation:}
We then focus our attention on the algorithmic question of optimal allocation 
among players with submodular valuations. 
We stress the difference between this case and the more restricted and 
well-studied case of
allocation among valuations satisfying the gross substitutes property.  
In the latter case,
it is known that a Walrasian equilibrium exists by a result of 
Kelso and Crawford(1982), while
in the former Walrasian equilibria do not necessarily exist: 
see Gul and Stacchetti (1999).  
Walrasian equilibria may be found
in polynomial time and are the basis of almost all known computationally 
efficient allocation algorithms.  
One may thus say that the submodular case is the first hard case 
from a computational point of view\footnote{A similar
phenomena exists from the information transfer point of view:
any optimal allocation algorithm among submodular valuations requires
exponential communication, while if the valuations satisfy the 
gross substitutes property then polynomial
communication suffices by a result of Nisan and Segal(2001).}.

We first show that the optimal allocation problem remains NP-hard
even among agents of submodular types.  
Our main positive algorithmic result is
a simple greedy algorithm that 
produces an allocation that is a $2$-approximation, i.e., an allocation
whose value is at least half of the optimal one.
This result is then generalized to types exhibiting marginal utilities that may
increase only in a bounded manner.
This is in sharp contrast to the general case where it remains NP-hard to find 
even a $n^{1/2-\epsilon}$-approximation: Sandholm(1999) and independently
Lehmann and al.(forthcoming).  

We do not know if a better approximation ratio is possible, 
nor do we know whether a polynomial time approximation algorithm exists 
that works for all complement-free valuations.

\item {\bf False Name Bids:}
Our next concern is with the strategic issue of {\em false-name bids} 
that was identified 
and analyzed in a sequence
of papers: Sakurai and al.(1999), 
Yokoo and al.(2000a, 2000b, 2001). 
This concerns the following problem that is inherent in combinatorial auctions 
that use the Vickrey-Clark-Groves payment rules 
(see MacKie-Mason and Varian(1994)) --
the only choice, if one requires incentive compatibility and efficiency.
Sakurai and al.(1999) observed that 
a bidder can manipulate a VCG combinatorial
auction and reduce his payment by splitting his bid and placing 
two separate bids under {\em false-names}.
They showed  that, 
in the case of single-item multi-unit auctions, 
this type of savings cannot occur when all bids are downward sloping.  
It is claimed there (without proof and without precise definitions) 
that this generalizes to combinatorial auctions 
when all bidders have no complementarities.
We show that this is not exactly the case\footnote{We have contacted the authors 
of this paper and they
have confirmed a bug in their unpublished proof.}, 
but rather the generalization to combinatorial auctions
requires that the {\em combined valuation} of other bidders is {\em submodular}.
\end{itemize}

We provide an example showing that it is not sufficient to
require that each player be of a submodular type.

\subsection{Paper Structure}
\label{sec:structure}
This paper is charting the so far unexplored territory of complement-free
valuations. It is therefore by necessity proceeding carefully and proving a
large number of small results.
In section~\ref{sec:defs} we present the basic definitions of the classes of 
valuations that we discuss.  
Section~\ref{sec:repandstruct} discusses the representation of valuations 
using OR and XOR expressions,
and provides a structural hierarchy of sub-classes of complement-free valuations.
It also shows that gap between submodular and gross substitutes valuations is
very large.  
Section~\ref{sec:examples} provides many examples of complement-free and 
submodular valuations, both natural ones. Such examples are used to separate
the different classes defined and studied previously.
Section~\ref{sec:allocation} presents the main result of this paper.
It discusses allocation (winner determination) algorithms for submodular agents,
comparing with the gross substitutes case.  
Section~\ref{sec:falsenames} discusses false-name bids. 
Finally, section~\ref{sec:bounded} shortly mentions an extension of the results 
of section~\ref{sec:allocation} to valuations exhibiting limited complementarities
and bounded increases in marginal utilities.

\section{Definitions}
\label{sec:defs}
\subsection{Preliminaries}
\label{sect:prelim}
In this paper we consider a combinatorial auction of non-identical goods.  
There is a set $X$ of items for sale by a
single auctioneer in a single combinatorial auction. 
We shall denote the number of items for sale by \mbox{m = $\mid X \mid$}.
There are $n$ agents (buyers, bidders) who all desire these items.  
Each agent $i$ has its own private valuation function, $v_i$, 
that specifies the utility it gets from each possible subset of items:  
i.e. for a subset \mbox{$A \subseteq X$} of items, 
$v_i(A)$ is the amount of money at which 
bidder $i$ values this subset $A$ of items.  
For a singleton set $\{x\}$ we will use
$v_i(x)$ as a shorthand for $v_i(\{x\})$.
This formalization makes two important implicit assumptions:
\begin{itemize}
\item
{\bf Quasi-linearity:} Agents' utilities can be measured in terms of ``money'', 
i.e. are linear in the ``money''.
\item
{\bf No Externalities:}  An agent's valuation depends only on
the set of items he wins, not on the identity of the agents 
that get the items outside $A$: the valuation function $v_i$ is
$v_i=v_i(A)$, where $A$ is the set of items won by agent $i$.
\end{itemize}
In addition, we assume that these valuations all
satisfy the following conditions:
\begin{itemize}
\item
{\bf Free disposal:} Items have non-negative value.  Thus $v_i$
satisfies $v_i(A) \le v_i(B)$, whenever \mbox{$A \subseteq B$}.
\item
{\bf Normalization:} $v_i(\emptyset)=0$.
\end{itemize}

\begin{definition}
\label{def:valuation}
A {\em valuation} is a function, \mbox{$v:2^X \rightarrow \cR^+$}, 
from subsets of items to the non-negative reals that
satisfies the normalization and free disposal conditions, 
i.e. \mbox{$v_i(\emptyset)=0$} and
\mbox{$v_i(A) \le v_i(B)$}, whenever \mbox{$A \subseteq B$} (monotonicity).
\end{definition}

The auctioneer's aim is to find an optimal {\em allocation}.

\begin{definition}
\label{def:allocation}
An {\em allocation} is a partition of $X$ into pairwise
disjoint sets of items $S_1 \ldots S_n$ that allocates the bundle $S_i$ to agent
$i$. An allocation is {\em optimal} if it maximizes
\mbox{$\sum_i v_i(S_i)$}.
\end{definition}
 
The auction rules must also define the payments received from each bidder.

\subsection{Marginal Valuations}
\label{sec:margin}
A central notion that we will be using is that of the marginal valuation.
It describes how a player would value sets of remaining items if he were already 
given some other items. 

\begin{definition}
\label{def:margval}
Given a valuation $v$ on a set $X$ of items and a set 
\mbox{$W \subseteq X$} of items, the {\em marginal valuation} of a set 
\mbox{$A \subseteq X-W$}
given $W$ is defined by:
\[
v(A \mid W) = v(A \cup W) - v(W).
\]
\end{definition}

One may consider the marginal valuation of a single element,
$v(x \mid \cdot)$, as a discrete analog for the partial derivative of
$v$ in direction $x$.

\subsection{No Complementarities}
\label{sec:over}
\begin{definition}
\label{def:CF}
A valuation $v$ is called {\em complement-free} if it satisfies:
\[
v(A) + v(B) \geq v(A \cup B)
\]
for all sets $A$, \mbox{$B \subseteq X$}.  
The class CF is the set of all complement free valuations.
\end{definition}

While this notion is clearly natural, it turns out that valuations 
with this property can still have {\em hidden complementarities}.
Example~\ref{xoscf} shows that even if $v$ is in CF, 
the valuation \mbox{$v(\cdot \mid W)$} need not be, i.e., 
once a set of items $W$ is already acquired, complementarities may surface.  
Indeed, as Example~\ref{xoscf} shows, this condition of complement-freedom
is not the proper analog of downward-sloping valuations.
It is therefore only natural to turn to those valuations 
that have no such hidden complementarities:
those valuations whose marginal valuations exhibit no complementarities. 
It turns out that these are exactly the submodular valuations, 
i.e., valuations with decreasing marginal utilities.

\subsection{Decreasing Marginal Utilities -- Submodular Valuations}
\label{sec:submodular}
This paper focuses on a subset of valuations that is of special 
economic interest: valuations with decreasing marginal utilities.  
These valuations are a strict subset of the complement-free valuations.
We believe that, in fact, it is the {\em correct} economic notion capturing 
{\em absolute lack of complementarity} and that it is a much more natural
assumption than the {\em convexity of preferences} that is so central to
much of economic theory. 
Such valuations are known as submodular functions.
In many combinatorial problems, submodularity can play a role similar to convexity
in optimization. 
\begin{definition}\label{dmu-def}
A valuation $v$ is called {\em submodular} if for
every two sets of items \mbox{$S \subseteq T$} and element $x$,
$v(x \mid T) \le v(x \mid S)$.  Submodular valuations are also called
valuations with {\em decreasing marginal utilities}.
The class SM is the set of all submodular
valuations.
\end{definition}
Thus we require that the marginal utility of an element decreases as the set
of items already acquired increases.  
Submodular functions are well-known
and heavily used in combinatorics (see Narayanan(1997).
If we extend the
analogy between the marginal valuation and the derivative discussed 
in section~\ref{sec:margin}, 
we see why submodular valuations are considered a
discrete analog of convex functions -- the ``derivative'' 
\mbox{$v(x \mid \cdot)$} is decreasing.
Here, we consider only valuations: monotone, positive submodular functions.
In the literature, submodular functions have been generally been considered
in a wider setting.
Many equivalent characterizations of decreasing marginal utilities
are well-known.

\begin{theorem}[see Moulin(1988) or Narayanan(1997)]
\label{the:submod}
A valuation $v$ is submodular if and only if any one of the following 
equivalent propositions holds.
\begin{itemize}
\item For any \mbox{$x, y \in X$} and \mbox{$S \subseteq X$}: 
$v(x \mid S) \geq v(x \mid S \cup \{y\})$.
\item For any \mbox{$S, T, V \subseteq X$}, such that
\mbox{$S \subseteq T$}: \mbox{$v(V \mid S) \geq v(V \mid T)$}.
\item \label{usualsubm} For any \mbox{$A, B \subseteq X$}:
\mbox{$v(A) + v(B) \geq v(A \cup B) + v(A \cap B)$}.
\end{itemize}
\end{theorem}

From the last characterization we get:
\begin{corollary}
A submodular valuation is complement-free.
\end{corollary}

The converse is not true and in Example~\ref{xoscf} below 
we exhibit a valuation that is complement-free, 
yet its marginal utilities are not decreasing.  
It turns out that valuations with decreasing
marginal utilities are exactly the valuations 
without any ``hidden'' complementarities.  

\begin{lemma}
\label{lemma:smmargin}[see Topkis(1998)]
A valuation $v$ is submodular if and only if for every
subset of items $R$, the marginal valuation function 
\mbox{$v(\cdot \mid R)$} is complement-free.
\end{lemma}

%

In particular we get the corollary.
\begin{corollary} 
\label{corr:res-cor}
If a valuation $v$ is submodular, then for every
subset $R$ of items the marginal valuation function 
\mbox{$v(\cdot \mid R)$} is submodular.
\end{corollary}
\begin{proof}
All the marginal valuation functions of \mbox{$v(\cdot \mid R)$} 
are also marginal valuation functions of $v$ and thus have no 
complementarities.
\end{proof}

\subsection{Gross Substitutes}
All the definitions and theorems in this subsection
are from Kelso and Crawford(1982), Gul and Stacchetti(1999).
We have sometimes slightly modified the terminology and notation.
We have kept with the now traditional terminology of {\em gross substitutes},
even though {\em gross} does not fit the situation.
In order to define the gross substitutes property 
we need to consider first the effect of putting prices on items. 
We think of what happens to the valuation of $v$ of a set $S$, when the price
of all items in $S$ must be paid.   

\begin{definition}
\label{def:surplus}
Given a vector of real item prices \mbox{$\vec{p} =$} \mbox{$(p_1 \ldots p_m)$}, 
the {\em surplus} of a set of items $S$
relative to these prices is defined as \mbox{$v(S \mid \vec{p}) =$} 
\mbox{$v(S) - \sum_{i \in S}p_i$}.  
A set $S$ is a {\em preferred set} of $v$ at prices 
\mbox{$\vec{p}$} if \mbox{$v(S \mid \vec{p}) =$} 
\mbox{$\max_T v(T \mid \vec{p})$}, i.e., $S$ maximizes the surplus. 
The {\em demand set} of $v$ at prices $\vec{p}$ is the set of
all preferred sets \mbox{$D(v \mid \vec{p}) =$} 
\mbox{$\{S\: \mid \:v(S \mid \vec{p}) = \max_T v(T \mid \vec{p})\}$}.
\end{definition}

The gross substitutes property mandates that increasing the price of an item will
not decrease the demand for any other item:  i.e., if an item $i$ is in a preferred
set at prices $\vec{p}$, then increasing $p_j$ for some $j \ne i$, 
will still have item $i$ in some preferred set.

\begin{definition}
\label{def:gross}
A valuation $v$ is said to satisfy the {\em gross substitutes} property 
if for any item $i$, any price vector 
$\vec{p}$ and any price vector $\vec{q} \geq \vec{p}$ 
(coordinatewise comparison) 
with $p_i = q_i$, we have that if $i \in A$ where 
\mbox{$A \in D(v \mid \vec{p})$}, 
then there exists \mbox{$A' \in D(v \mid \vec{q})$} such that $i \in A'$.  
The class GS is set of all valuations that satisfy the gross substitutes 
property.
\end{definition}

An equivalent condition is called the {\em single improvement property}.  
This condition states that a non-preferred set can always be improved 
(in terms of its surplus) by deleting at most one element
from it and inserting at most one element into it.

\begin{lemma}[Gul and Stacchetti(1999)]
\label{si} 
A valuation $v$ satisfies the gross substitutes property if and only if 
for any $\vec{p}$ and $A \not\in D(v \mid \vec{p})$,
there exists $A'$ such that 
\mbox{$ \mid A' - A \mid  \le 1$} and 
\mbox{$ \mid A - A' \mid  \le 1$} and 
\mbox{$v(A' \mid \vec{p}) > v(A \mid \vec{p})$}.
\end{lemma}

The gross substitutes property is stronger than submodularity.

\begin{lemma}[Gul and Stacchetti(1999)]
A valuation that satisfies the gross substitutes property is submodular.
\end{lemma}

The key property that makes valuations with the gross substitutes property 
so convenient is that in an
auction (equivalently an exchange economy) with such valuations, 
Walrasian equilibria exist.  
A Walrasian equilibrium is a vector of prices
on the items and an allocation 
such that every bidder receives a preferred set
at these prices.

\begin{definition}
\label{wal}
A Walrasian equilibrium in an auction with valuations 
\mbox{$v_1 \ldots v_n$} is a price vector $\vec{p}$
and an allocation \mbox{$A_1 \ldots A_n$} 
such that for all bidders $j$: 
\mbox{$A_j \in D(v_j \mid \vec{p})$}.
\end{definition}

\begin{theorem}[Kelso and Crawford(1982)]
\label{gswal}  
Any auction with gross substitutes valuations $v_1 \ldots v_n$ 
has a Walrasian equilibrium.
\end{theorem}

It turns out that this is essentially also a necessary condition.

\begin{theorem}[Gul and Stacchetti(1999)] 
\label{ngsnwal} 
If $v_1$ does {\em not} satisfy the gross substitutes property 
then there exist valuations $v_2 \ldots v_n$ that do satisfy the gross substitutes 
property\footnote{In fact, Gul and Stacchetti(1999) prove that
$v_2 \ldots v_n$ may be unit demand functions -- a stronger claim.} 
and the auction with $v_1 \ldots v_n$ does {\em not} have a 
Walrasian equilibrium.
\end{theorem}

It is easy to verify the ``first Welfare theorem'' in this context: 
any Walrasian equilibrium gives an optimal allocation, i.e. one that
maximizes $\sum_j v_j(A_j)$.  

\section{Representation and Structure}
\label{sec:repandstruct}
\subsection{Elements of Representation}
The question of representation of valuations, i.e., 
how to succinctly describe a valuation without listing explicitly 
the values for each of the $2^n-1$ subsets of items,
must be addressed before valuations can be efficiently treated 
in any allocation algorithm.
How can we represent complement-free valuations succinctly?
We may consider representing such valuations by combining 
{\em simple} valuations by suitable operators.

The basic syntactic elements from which valuations are usually constructed 
consist of a certain price for a specific set of elements.
They correspond to the single-minded valuations (i.e., bidders) of 
Lehmann and al.(forthcoming).
The only single-minded valuations in CF
are valuations in which only a single item is valued at a positive value.
It is therefore expected that those {\em singleton} valuations 
will play a central role in describing CF valuations.

\begin{definition}
\label{def:single}
For an item $x \in X$ and a price $p$, 
the singleton valuation $e_{x}^{p}$ is a valuation giving the same value $p$ 
to all sets containing $x$ and the value $0$ to all sets that
do not include $x$.
\end{definition}

The operators commonly used for representing valuations are 
those introduced in Sandholm(1999): OR and XOR.
They are often understood as prescriptions concerning the compatibility of
bids, but the following presentation, as operations on valuations, is more
convenient.

\begin{definition}[Nisan(2000)] 
\label{def:ORXOR}
Let $v_1$ and $v_2$ be two valuations on the set $X$ of items.
The valuations \mbox{$v_1 \oplus v_2$} (XOR) and the valuation 
\mbox{$v_1 \vee v_2$} (OR) are defined by:
$$ (v_1 \oplus v_2)(S) = max( v_1(S), v_2(S) ),$$
$$ (v_1 \vee v_2)(S) = \max_{T \subseteq S}{(v_1(T) + v_2(S-T))}. $$
\end{definition}


Both operators have an intuitive description.
The OR ($\vee$) of two valuations represents the valuation 
of an agent representing both valuations,
bidding on their behalf and sharing the result between them optimally. 
The value given to a bundle by the OR of two valuations 
is obtained by partitioning the bundle optimally between these valuations. 
The XOR ($\oplus$) of two valuations represents the valuation of a single bidder 
capable of choosing between two possible but incompatible personalities 
after the auction.
His value for a set $S$ is the largest of the two values 
of the component valuations.
The operations OR and XOR are obviously associative and commutative.

\subsection{A Syntactic Hierarchy}
\label{syntax}
It is now natural to study the hierarchy of valuations obtained by
combining singleton valuations by the OR and XOR operations.  
Here are the levels of the hierarchy

\begin{enumerate}
\item
Singleton valuations.
\item
$OS-valuations$:  
The family of valuations that can be described by OR of singletons 
is precisely the family of (separately) {\em additive valuations}: 
those valuations which value any set at the sum of the values of its elements.
They exhibit no complementarities and no substitutabilities.
\item $XS-valuations$: 
The family of valuations that can be described by XOR of singletons
is usually called the family of {\em unit-demand valuations} 
Gul and Stacchetti(1999): those valuations that value any set
of elements at the value of the element of the set that is most valued.
This family of types has been considered central by Vickrey and since.
\item
$OXS-valuations$: 
The family of OR's of XS valuations represents the aggregated valuation
of multiple unit-demand players.
\item
$XOS-valuations$: 
The family of XOR's of additive valuations turns out to subsume the 
whole hierarchy. 
This follows from the simple observation that OR distributes over XOR:
\[
(a \oplus b) \vee (c \oplus d) = 
(a \vee c) \oplus (a \vee d) \oplus (b \vee c) \oplus (b \vee d).
\]
Therefore XOS is closed under OR (it is obviously closed under XOR).  
It follows that any expression
that uses combinations of OR and XOR operations on singleton valuations
can be represented as a simple XOS expression.
\end{enumerate}

We shall see in Section~\ref{subsec:sep} that XOS is a strict subset of CF, 
but contains SM  
and that OXS is a strict subset of XOS.
In fact we have just shown that on any set of atomic valuations 
(not only singletons), 
XOR-of-OR combinations can represent arbitrary complex 
combinations of ORs and XORs.

\subsection{Closure Theorems}
Once we have defined the operations of OR and XOR it is natural to ask whether
the classes of valuations defined semantically (CF, SM, and GS) are closed under these operations.

\subsubsection{CF}
The operations of OR and XOR fit well our concern with the class CF since:
\begin{theorem}
\label{the:CFclosed}
The class CF is closed under OR and XOR.
\end{theorem}
\begin{proof} 
Let $v_1$ and $v_2$ be complement free valuations then:
\[
(v_1 \oplus v_2 )(S \cup T) = max(v_1 (S \cup T), v_2 (S \cup T)) \le
\]
\[
max(v_1 (S) + v_1 (T) , v_2 (S) + v_2 (T)) \le
\]
\[
max(v_1 (S) , v_2 (S)) + max(v_1 (T) , v_2 (T)) =
\]
\[
(v_1 \oplus v_2 )(S) + (v_1 \oplus v_2) (T).
\]

Similarly fix partitions $(S_1 : S_2)$ of $S$ and $(T_1 : T_2)$ of $T$,
such that
\mbox{$(v_1 \vee v_2 )(S \cup T) =$}
\mbox{$v_1(S_1 \cup T_1) + v_2(S_2 \cup T_2)$}.
Now,
\[
(v_1 \vee v_2 )(S \cup T) = v_1 (S_1 \cup T_1) + v_2 (S_2 \cup T_2) \le
\]
\[
v_1 (S_1) + v_1 (T_1) + v_2 (S_2) + v_2 (T_2)) \le
\]
\[
(v_1 \vee v_2 )(S) + (v_1 \vee v_2) (T).
\]
\end{proof}

\subsubsection{SM}
The class of submodular valuations is {\em not} closed under either OR or XOR.
It is not closed under XOR as it contains
the class OS of additive valuations, 
but Example~\ref{smxos} shows that it does not contain
the class XOS which is XORs of additive valuations.  
The fact that it is not closed under OR
is shown by Example~\ref{ex:OR}.

\subsubsection{GS}
The class GS is not closed under XOR as it contains
the class OS of additive valuations, 
but Example~\ref{smxos} shows that even its superset SM does not contain
the class XOS which is XORs of additive valuations.  
It is however closed under the OR operation, as will be shown now.  

\begin{theorem}
\label{OGS}
If valuations $u$ and $w$ satisfy the gross substitutes property 
then so does $u \vee w$.
\end{theorem}

The theorem follows from combining the results of Fujishige and Yang(2001) 
showing that gross substitutes is equivalent to $M^\#-concavity$ with those of
Murota and Shioura(1999) regarding the closure of $M^\#$-convex functions under
convolution (i.e., OR).  
For completeness, we provide a simple proof based on the results cited above of 
Kelso and Crawford(1982) and Gul and Stacchetti(1999).

\begin{proof}
We will prove the theorem indirectly, applying Theorems~\ref{ngsnwal} 
and~\ref{gswal}.  
Using Theorem~\ref{ngsnwal}, it suffices to show that for all gross substitutes
valuations \mbox{$v_2 \ldots v_n$}, an auction among agents 
\mbox{$u \vee w, v_2 \ldots v_n$} has a Walrasian equilibrium.  
Theorem~\ref{gswal} states that an auction
among agents \mbox{$u, w, v_2, \ldots , v_n$} has a Walrasian equilibrium: 
vector of prices $\vec{p}$ with allocation \mbox{$A_u, A_w, A_2, \ldots A_n$}.  
We claim that  $\vec{p}$ with the allocation
\mbox{$A_u \cup A_w, A_2, \ldots A_n$} is a Walrasian equilibrium for bidders 
\mbox{$u \vee w, v_2 \ldots v_n$}.

The only thing that needs to be verified is that 
\mbox{$(A_u \cup A_w) \in D(u \vee w \mid \vec{p})$}.  
This is so since for any set $S$, let $S_u$ and $S_w$ be a partition of $S$ 
such that
\mbox{$(u \vee w)(S) =$} \mbox{$u(S_u) + w(S_w)$}.  
Thus also
\mbox{$(u \vee w)(S  \mid  \vec{p}) =$} 
\mbox{$u(S_u \mid \vec{p}) + w(S_w \mid \vec{p})$}.
Now, since \mbox{$A_u \in D(u \mid \vec{p})$} we have that
\mbox{$u(S_u \mid \vec{p}) \le$} \mbox{$u(A_u \mid \vec{p})$}, 
and similarly for $w$. 
Clearly \mbox{$u(A_u) + w(A_w) \le$} \mbox{$(u \vee w)(A_u \cup A_w)$}, 
and so also
\mbox{$u(A_u \mid \vec{p}) + w(A_w \mid \vec{p}) \le$} 
\mbox{$(u \vee w)(A_u \cup A_w \mid \vec{p})$}.
Thus
\[
(u \vee w)(S  \mid  \vec{p}) = u(S_u \mid \vec{p}) + w(S_w \mid \vec{p}) \le \]
\[
\le u(A_u  \mid  \vec{p}) + w(A_w  \mid  \vec{p}) \le (u \vee w)(A_u \cup A_w  \mid  \vec{p}).
\] 
\end{proof}

\subsection{The Complete Hierarchy}
\label{sec:hierarchy}
We have seen by now five significant classes of valuations.  
Three of them were defined semantically:
complement-free (CF), submodular (SM), and gross substitutes (GS).  
Two of them were
defined syntactically: ORs of XORs of singletons (OXS) and XORs of ORs 
of singletons (XOS).  We now can state the relationships between these classes.

\begin{theorem}
\label{the:hierarchy}
\[
OXS \subset GS \subset SM \subset XOS \subset CF.
\]
All containments are strict.
\end{theorem}
\begin{proof}
We will prove the containments from left to right.  
The fact that they are strict will follow from the examples found below 
in Section~\ref{subsec:sep}.
\begin{enumerate}
\item $OXS \subseteq GS$: 
This follows from Theorem~\ref{OGS} since GS 
contains all unit demand valuations  (XS-valuations) (Gul and Stacchetti(1999)), 
and therefore
also contains ORs of XS valuations.
\item $GS \subseteq SM$: 
This was shown in Gul and Stacchetti(1999).
\item $SM \subseteq XOS$: 
each submodular valuation may be represented 
by a long XOR expression.
For each permutation $\pi$ of the items we will have an OR clause -- 
the XOS expression will be the XOR of all these OR clauses.  
The OR clause for the permutation $\pi$ will offer for each item $i$ 
the marginal price of $i$ 
assuming that all items preceding it in the permutation $\pi$ 
have already been obtained.
Formally, for $i = \pi(j)$, the price is 
\mbox{$v(i \mid \{\pi(1), \ldots ,\pi(j-1)\})$}. 
To see that the XOS expression indeed represents the submodular valuation, 
consider a submodular $v$ and
a set $A$ of items. For any permutation $\pi$ of the items 
in which the items of $A$ are placed first,
the value given to $A$ by the OR clause for $\pi$ is exactly $v(A)$.
All other OR clauses give to $A$ a value that is smaller or equal,
since $v$ has decreasing marginal utilities.
\item $XOS \subseteq CF$:
By Theorem~\ref{the:CFclosed}.
\end{enumerate}
\end{proof}

\subsection{Measure}
In this section we study how $large$ are the five classes of valuations 
studied so far.  
Our sense of size is the simple one of measure.  
Each valuation is specified by a vector of $2^m-1$ real numbers, and
we can thus look at each class of valuations as residing 
in the $2^m-1$-dimensional Euclidean space.  
What we ask here is what is the measure of these classes, 
in particular whether their measure is positive. 
It turns out that there is a significant gap between the GS and SM levels of 
the hierarchy: GS valuations have measure $0$,
while SM valuations have positive measure.

\begin{theorem}
For every number of items $m \ge 3$, the class GS (on $m$ items) 
has zero-measure in the $2^m-1$-dimensional Euclidean space.  
Therefore its interior is empty.
\end{theorem}

\begin{proof}
The theorem is directly implied by the following claim:
\begin{claim}
\label{the:interior}
If \mbox{$x , y , z \in \Omega$} are pairwise distinct items and $v$ is a GS 
valuation such that 
\mbox{$v(z \mid y) < v(z \mid x)$}, then
\mbox{$v(x \mid y) = v(x \mid z)$}.
\end{claim}

The theorem follows from the claim since the set GS is contained in the union of 
the three hyper-planes defined by the equations 
\mbox{$v(x \mid y) = v(x \mid z)$} for all permutations of $x, y, z$.

The claim follows from the two lemmas below. 
Each of those lemmas is proved by a
similar technique, making a direct use of a GS valuation: the prices of two of
the three items are fixed and the price of the third item is made to increase from
zero to a large value.
\begin{lemma}
\label{le:intone}
If \mbox{$v(z \mid y) < v(z \mid x)$}, then \mbox{$v(x \mid y) \geq v(x \mid z)$}.
\end{lemma}
\begin{proof}
Given a valuation $v$, a price vector (for all items), 
and sets $A$, $B$ of items, we write
\mbox{$A \preceq B$} or \mbox{$B \succeq A$} if $B$ is at least 
as preferred as $A$ at the given prices,
\mbox{$A \prec B$} or \mbox{$B \succ A$} if $B$ is strictly preferred to $A$ and
\mbox{$A \sim B$} if $A$ and $B$ have the same net utility given the prices.
Let us fix the prices of $x$ and $y$:
\mbox{$p_{x} = v(x \mid z)$} and \mbox{$p_{y} = v(y \mid z)$}.
These prices induce preferences between subsets of $\Omega$.
First:
\mbox{$xz \sim z \sim yz$}. 
Notice that this holds independently of the price of $z$.
Notice also that, since $v$ is GS it is submodular, i.e., 
it exhibits decreasing marginal values and, 
since one does not gain by adding $x$ to $z$, one cannot gain 
by adding $x$ to $yz$ and therefore
\mbox{$xyz \preceq yz$}.
Let us now fix, at first, the price of $z$ at zero: \mbox{$p^{1}_{z} = 0$}.
One never loses by adding $z$ to one's bundle and therefore all three bundles:
$z$, $xz$ and $yz$ are preferred bundles. 
In particular $xz$ is a preferred set at the current prices. 
Let us now increase the price of $z$ to some value,
\mbox{$p^{2}_{z}$} such that
\mbox{$v(z \mid y) < p^{2}_{z} < v(z \mid x)$}.
At the new prices \mbox{$yz \prec y$} and \mbox{$x \prec xz$}. 
Therefore \mbox{$x \prec y$}. 
By the gross substitutes property, there is, at the new prices, some
preferred set that contains $x$. This set must be at least as preferred as $y$ 
and therefore cannot be $x$ and certainly not $xz$ or $xyz$. 
Therefore $xy$ is a preferred set and \mbox{$xy \succeq y$}:
\mbox{$v(x \mid y) - p_{x} \geq 0$} and 
\mbox{$v(x \mid y) \geq v(x \mid z)$}.
\end{proof}
\begin{lemma}
\label{le:inttwo}
If \mbox{$v(z \mid y) < v(z \mid x)$}, 
then \mbox{$v(x \mid y) \leq v(x \mid z)$}.
\end{lemma}
\begin{proof}
By contradiction. Assume \mbox{$v(z \mid y) < v(z \mid x)$} and 
\mbox{$v(x \mid y) >$} \mbox{$v(x \mid z)$} 
and fix prices for $x$ and $z$ such that:
\mbox{$v(x \mid z) <$} \mbox{$p_{x} <$} \mbox{$v(x \mid y)$} and
\mbox{$v(z \mid y) <$} \mbox{$p_{z} <$} \mbox{$v(z \mid x)$}.
Those prices for $x$ and $z$ imply the following preferences:
\mbox{$yz \prec y \prec xy$} and \mbox{$x \prec xz \prec z$}.
Since $v$ is submodular, we also know that \mbox{$xyz \prec yz$} 
(one does not gain by adding $x$ to $z$, therefore one cannot gain by adding
$x$ to $yz$) and that \mbox{$\emptyset \prec x$} (one gains by adding $x$ to $y$,
therefore one gains by adding $x$ to the empty bundle).
Let us fix, first the price of $y$ at zero. Then, $xy$ is a preferred set.
Let, then, the price of $y$ increase to a value high enough to make any bundle 
containing $y$ unattractive. Then $z$ is the only preferred bundle, contradicting
the substitutes property.  
This concludes the proof of the lemma, of the claim, and of the theorem.
\end{proof}
\end{proof}

We do not know what is the dimensionality of the class GS.  
In particular it is possible that its dimension is only polynomial in $m$.
We have shown that the class GS is {\em small}, let us now show that SM is 
{\em large}.
\begin{theorem}
\label{the:large}
For any number of items $m\ge 1$, the class SM of valuations over $m$ items
has a non-empty interior in the $2^m-1$-dimensional Euclidean space.  
Therefore it has positive measure.
\end{theorem}

\begin{proof}
We shall use \mbox{$\mid x \mid$} to denote
the absolute value of a number $x$ and \mbox{$\mid S \mid$} 
to denote the cardinality of a finite set $S$.
Let \mbox{$n = \mid \Omega \mid$}.
The valuation \mbox{$v_{c}(S) = 1 - 2^{- \mid S \mid}$} 
is a symmetric (see section~\ref{subsec:symmetric}), monotone, normal, 
submodular valuation for which the marginal value of the $k$th element is $2^{-k}$.
It is, in a sense, a {\em central} submodular valuation.
The theorem follows from the claim that, for any function 
\mbox{$h : 2^{\Omega} \rightarrow \cR$} such that
\mbox{$\mid h(S) \mid \leq 2^{-(n+2)}$}, the valuation 
\mbox{$v(S) = v_{c}(S) + h(S)$} is submodular.
Indeed, assume \mbox{$S \subset T$} and \mbox{$x \not \in T$}. We have:
\[
v(x \mid S) = v(S \cup \{x\}) - v(S) \geq 2^{-(\mid S \mid + 1)} - 2^{-(n+1)} \geq
\]
\[
2^{-(\mid S' \mid + 1)} + 2^{-(\mid S' \mid + 1)} - 2^{-(n+1)}  \geq
2^{-(\mid S' \mid + 1)} + 2^{-(n+1)} \geq
\]
\[
v(x \cup S') - v(S) = v(x \mid S').
\] 
\end{proof}

\section{Examples of Submodular Valuations}
\label{sec:examples}
\subsection{Natural Examples}
The simplest and most common natural examples of valuations 
with no complementarities
are the additive valuations
(the class OS) and the unit demand valuations (the class XS) mentioned in
Section~\ref{syntax}.  
These valuations are both trivially in sub-classes of OXS and thus clearly
satisfy the gross substitutes property, and are also submodular.
We now mention some other natural valuations that are submodular.

\subsubsection{Symmetric valuations}
\label{subsec:symmetric}
Symmetric valuations are ones where $v(S)$ depends only on the size of $S$, 
$\mid S \mid$.   
These cases fit auctions of multiple identical items. 
Symmetric valuations may be described by a sequence of {\em marginal} values:
numbers \mbox{$p_1 \ldots p_m$} 
where \mbox{$p_{i} = v(T) - v(S)$} for any sets $T$ and $S$ of sizes
$i$ and $i-1$ respectively.
We then have: \mbox{$v(S) = \sum_i^{ \mid S \mid } p_i$} for any set $S$.  
The following set of symmetric valuations has received considerable attention.  
For example Vickrey(1961)'s multi-unit auction considers only such valuations.

\begin{definition}
\label{def:downslope}
A symmetric valuation is downward sloping iff for all $i$, $p_{i+1} \le p_{i}$.
\end{definition} 
As expected, these are exactly the symmetric submodular valuations.
\begin{proposition}
\label{prop:down}
A symmetric valuation is downward sloping if and only if it is submodular.
\end{proposition}
\begin{proof}
For a symmetric valuation $v$, for \mbox{$x \not \in S$}, 
$v(x \mid S)=p_{ \mid S \mid +1}$, and the equivalence is immediate.
\end{proof}
It is known from Gul and Stacchetti(1999) that downward sloping 
symmetric valuations also satisfy
the gross substitutes property and in fact
can be represented in 
the class OXS (see Nisan(2000)). 
On the other hand, the complement-free symmetric valuations are a wider
class -- see below Examples~\ref{xoscf} and~\ref{smxos}.

\subsubsection{Additive valuations with a budget limit}
\label{subsec:budgetl}
Our second class of examples is composed of valuations that are 
additive-up-to-a-budget-limit.  
In these valuations, a set is valued at the sum of the values of its items, 
unless this sum is larger than a budget limit. 
In this last case it is valued at the budget limit.
\begin{definition}
\label{def:abudgetl}
A valuation $v$ is called additive with a budget limit 
if there exists a constant $b$, the budget limit,
such that for all sets $S$ of items, \mbox{$v(S) =$} 
\mbox{$min(b, \sum_{i \in S} v(i) )$}.
\end{definition}
\begin{proposition}
\label{prop:budget}
Every additive valuation with a budget limit is submodular.
\end{proposition}
\begin{proof}
Assume \mbox{$S \subseteq T$} 
and \mbox{$x \not \in T$}.
We have 
\[
v(x \mid T) = min(b, v(T) + v(\{ x \})) - v(T) = 
\]
\[
min(b-v(T), v(x)) \leq min(b-v(S), v(x)) = v(x \mid S).
\]
\end{proof}
These type of valuations do not necessarily satisfy 
the gross substitutes property -- see below Example~\ref{gssm}.

\subsubsection{Valuations based an underlying measure}
The third class of valuations we want to present consists of valuations 
based on some underlying set with a measure on it.
Let us first start with a concrete example.  
Consider a combinatorial auction for a set of spectrum licenses 
in overlapping geographical regions.  
Each geographical region contains a certain population, 
and a reasonable valuation for a set of licenses is the total population 
in the geographic area covered by the regions in the set.  
It turns out that such valuations are submodular.  
More generally, the overlapping regions may be arbitrary sets 
in some underlying base set, 
and the population count may be replaced by any measure on the base set.
\begin{definition}
\label{def:underl}
A valuation $v$ is said to be based on the underlying measure $\mu$ 
on a base set $\Gamma$ if there are
sets $I_{1} \ldots I_{m} \subseteq \Gamma$ such that for each set $S$ of goods, 
$v(S) = \mu ( \cup_{i \in S} I_{i} )$.
\end{definition}
\begin{lemma}
Every valuation that is based on an underlying measure is submodular.
\end{lemma}
\begin{proof}
Consider \mbox{$S \subseteq T$}, and an item $x$.  
Denote \mbox{$\tilde{S} =$} \mbox{$\cup_{i \in S} I_{i}$}, and 
similarly \mbox{$\tilde{T}$}.  Clearly \mbox{$\tilde{S} \subseteq \tilde{T}$}. 
We now have \mbox{$v(x \mid S) =$} 
\mbox{$\mu(I_{x} - \tilde{S}) \geq \mu(I_{x} - \tilde{T}) =$}
\mbox{$v(x \mid T)$}.
\end{proof}

\subsection{Separation Examples}
\label{subsec:sep}
\subsubsection{$OXS \ne GS$}
\begin{example}
Consider a set $X$ of four items: \mbox{$X = \{a , b , c , d\}$}
and the valuation $v$ defined by: 
the value of any singleton is $10$
and the value of any other set is $19$, except for the two sets 
\mbox{$\{a , c\}$} and \mbox{$\{b , d\}$} the value of which is $15$.
\end{example}

\begin{claim}
\label{notOXS}
The valuation $v$ is not in OXS.
\end{claim}
\begin{proof}
Suppose $v = v_{1} \vee \ldots \vee v_{n}$.
Since \mbox{$v(\{a\})=10$}, there is some $i$ with $v_{i}(\{a\}) = 10$.
Without loss of generality, assume \mbox{$i = 1$}.
Since \mbox{$v(\{b\})=10$}, there is some $i$ with $v_{i}(\{b\}) = 10$.
Since \mbox{$v(\{a , b\}) < 20$}, for any \mbox{$i \neq 1$}, 
\mbox{$v_{i} < 10$}. We conclude that \mbox{$v_{1}(\{b\}) = 10$}.
Similarly, $v_{1}(\{c\}) = v_{1}(\{d\}) = 10$.
Consider now that \mbox{$v(\{a , c\}) = 15$}, therefore, 
since $v_{1}(\{a\}) = 10$, for any 
\mbox{$i > 1$}, \mbox{$v_{i}(\{c\}) \leq 5$}.
Similarly, for any item $x$ and any 
\mbox{$i > 1$}, \mbox{$v_{i}(\{x\}) \leq 5$}.
But \mbox{$v(\{a , b\}) = 19$} and therefore there must be \mbox{$i \neq j$}
such that \mbox{$v_{i}(\{a\}) + v_{j}(\{b\}) = 19$}, a contradiction.
\end{proof}

\begin{claim}
\label{GS}
The valuation $v$ is in GS.
\end{claim}
\begin{proof}(sketch)
We will use the single improvement property (Lemma \ref{si}).  
Let $\vec{p}$ be any
price vector and consider any non-preferred set $A$.  Let $D \in D(v \mid \vec{p})$
be a preferred set.  We need to show that we can improve 
the surplus of $A$ by deleting at most one element 
and inserting at most one element. 
This requires a tedious case by case analysis that we omit, except for the only nontrivial case where
$A=D^c$ and both are of size 2, e.g. $A=\{a,b\}$ and $D=\{c,d\}$.  In this case 
either $\{a,d\}$ or $\{b,c\}$ will have a surplus greater than $A$'s.
\end{proof}

\subsubsection{$GS \ne SM$}
\begin{example}\label{gssm}
Consider the additive valuation with budget limit 4 on three elements:
$v({1}) = v({2}) = 2$, $v({3}) = 4$. Since the budget limit is 4, 
each set that contains at least two elements 
has a valuation of 4.
\end{example}

This valuation is submodular by Proposition~\ref{prop:budget}, 
but it does not satisfy the gross substitutes property.
Consider the price vector $p_1=0, p_2=1, p_3=2$.  At these prices the preferred
subset is $\{1,2\}$ giving a surplus of 3.  However, if we increase the
price of item 1 to $p_1=2$, then the preferred subset is $\{3\}$ 
giving a surplus of
$2$, where any set that contains item 2 will have a surplus of at most 1.  
Thus we have reduced the demand for item 2.

\subsubsection{$SM \ne XOS$}
\begin{example}\label{smxos}
Consider the symmetric valuation on three elements defined by
\mbox{$p_1 = 2$}, \mbox{$p_2 = 0$}, \mbox{$p_3 = 1$}: a set of
one or two elements is valued at 2, while the set of all three
elements is valued at 3.  
\end{example}
By definition, it is not downward sloping, therefore not submodular,
by Proposition~\ref{prop:down}, but it is obtained by the following
XOR-of-OR-of-singletons: 
\[
(\{1\}:2) \oplus (\{2\}:2) \oplus (\{3\}:2) \oplus w
\]
where \mbox{$w =$} \mbox{$(\{1\}:1) \vee (\{2\}:1) \vee (\{3\}:1)$}.

\subsubsection{$XOS \ne CF$}
\begin{example}\label{xoscf}
\label{ex:CFnotclosedumarg}
Assume a set $X$ of three items and \mbox{$v(S) = 2$} if \mbox{$\mid S \mid = 1$},
\mbox{$v(S) = 3$} if \mbox{$\mid S \mid = 2$} and 
\mbox{$v(S) = 5$} if \mbox{$\mid S \mid = 3$}.
\end{example}

The valuation $v$ is in CF since \mbox{$2 + 2 \geq 3$} and \mbox{$2 + 3 \geq 5$}.  
We shall show that it cannot be expressed as a XOR-of-ORs-of-singletons.  
Assume that it is, and consider the OR clause that provides the valuation of $5$ 
to the set of all three elements.  
This OR clause contains (at most) three singleton
bids for the three items; 
if we take the two highest bids of these three we must get a valuation of at
least $5 \cdot 2/3 > 3$, 
in contradiction to the valuation of any two elements being 3.

This is also an example for the fact that the class CF is not closed 
under marginal valuations.
The marginal valuation $v_{W}$, where $W$ contains any single item, gives
\mbox{$v_{W}(S) = 1$} if \mbox{$\mid S \mid = 1$} and
\mbox{$v_{W}(S) = 3$} if \mbox{$\mid S \mid = 2$} and is not in CF.

\subsubsection{SM is not closed under OR}
The following example will show that SM is not closed under OR.
\begin{example}
\label{ex:OR}
Let $u(1) = 3$, $u(2) = 5$ and $u(3) = 3$, where $u$ has a budget limit of 6.
Let $w$ be the additive valuation: \mbox{$w(1) = 1$}, \mbox{$w(2) = 2$}, 
\mbox{$w(3) = 0$}.
Let \mbox{$v = u \vee w$}.
\end{example}
The valuation $u$ is submodular by Proposition~\ref{prop:budget}.
One can see that: \mbox{$v(\{1,2\}) =$} \mbox{$6$} ($u$ gets both), 
\mbox{$v(\{2,3\}) = 6$} 
($u$ gets both).
Therefore \mbox{$v(\{1,2\}) + v(\{2,3\}) =$} \mbox{$12$}.
But $v(\{1,2,3\}) = 8$ ($u$ gets 1 and 3 and $w$ gets 2), 
\mbox{$v(2) = 5$} ($u$ gets it), and \mbox{$8 + 5 > 12$}
and $v$ is not submodular.

We know of two special cases in which the OR of submodular functions is submodular.
The first concerns symmetric valuations.
\begin{proposition}
\label{prop:symmOR}
If $v_1$ and $v_2$ are symmetric submodular valuations then 
\mbox{$v_1 \vee v_2$} is submodular
(and symmetric).
\end{proposition}
\begin{proof}
By proposition~\ref{prop:down} and Nisan(2000), 
any symmetric submodular valuation is an 
OXS valuation. Since OXS is clearly closed under the $OR$ operation we have that
\mbox{$v_1 \vee v_2$} is in OXS and thus also, by Theorem~\ref{the:hierarchy}, 
is submodular. 
\end{proof}
The second case concerns valuations with disjoint supports.
\begin{proposition}
\label{prop:disjOR}
If $v_1$ and $v_2$ are submodular valuations with disjoint supports 
(i.e., \mbox{$v_{i}(x) > 0$} implies
\mbox{$v_{j}(x) = 0$}) then \mbox{$v_1 \vee v_2$} is submodular.
\end{proposition}
\begin{proof}
For any set $S$: \mbox{$(v_1 \vee v_2)(S) = v_1(S) + v_2(S)$}.
\end{proof}

\section{Allocation}
\label{sec:allocation}
We now turn to the computational problem of allocating the items 
in a combinatorial auction in which all agents have submodular valuations. 
As a computational problem, we must first consider the
format of the input, i.e. how are the valuation functions $v_1 \ldots v_n$ 
presented to the algorithm.
In the most general case representing a valuation may require exponential size 
(in $m$, the number of items).  
We, on the other hand, are looking for algorithms that are polynomial 
in the relevant parameters, $n$ and $m$.  
There are two possible approaches for obtaining efficient algorithms 
despite the exponential size of the input.  

The first approach considers the general case, but where the
valuations are presented to the algorithm as ``valuation oracles'' 
-- black boxes that can be queried for the valuation of a set $S$, 
returning its valuation $v(S)$.  
In a mechanism, this corresponds to allowing
bidders to send an arbitrary representation of the valuation function, 
as long as the valuations of sets
can be efficiently computed from it\footnote{Surprisingly, achieving this is 
not totally trivial -- see section 3.6 of Nisan(2000).}.  
The second approach fixes a representation, 
and provides allocation algorithms that require
polynomial time in the size of the representation of the valuations 
in this format.  
Such algorithms will generally be as interesting as the strength
of the representation format.  

We present our main positive result in section~\ref{subsec:2approx}, 
in the general terms of valuation oracles.  
Our main negative result, the NP-hardness of allocation for
submodular valuations, presented in subsection~\ref{subsec:hard}, 
holds for a special case (additive valuations with a budget limit) 
that has a simple short representation. Both our results are therefore
presented in the strongest possible manner.
Concrete lower bounds for the valuation oracle representation are derived 
in Nisan and Segal(2001).  
We first review the case of GS agents, which is polynomial time solvable.

\subsection{The Case of Gross Substitutes}

The case of a combinatorial auction where all valuations satisfy 
the gross substitutes property can be solved in polynomial time.  
Since Walrasian equilibria exist in this case by 
Kelso and Crawford(1982) (see definition~\ref{wal} 
and theorem~\ref{gswal} above), 
they can be found and an optimal allocation results.  
Let us look more closely at the computational process of finding 
this Walrasian equilibrium.
The following procedure of Kelso and Crawford(1982) yields prices and an
allocation that is arbitrarily close to optimal: 
\begin{enumerate}
\item Initialize all item prices to 0. 
\item
Repeat the following
procedure: compute
a preferred set for each bidder at these prices and increase 
by a small amount $\epsilon$ 
the price of all items that are demanded
by more than one agent.  
\item Stop when each item is demanded by at most one agent, and let the allocation
be the preferred sets at these prices.  
\end{enumerate}
This procedure produces an allocation whose distance from optimal
depends only on $\epsilon$ and whose running time is polynomial in 
\mbox{$n, m, \epsilon^{-1}$}.  
This algorithm is naturally implemented as a mechanism, 
in some cases is actually incentive compatible, and in the general case can
serve as a basis for an incentive compatible mechanism as in Ausubel(2000).
But this procedure only provides an approximation (it is actually a ``FPAS'' --
fully polynomial approximation scheme) and not an exact solution.
  
An exact solution can be computed in polynomial time using linear programming. 
Indeed, the optimization problem can be described 
as an integer-programming problem whose LP relaxation has an integral solution 
(see Bikhchandani and al.(2001)).
This linear program has an exponential number of variables, 
but only a polynomial number of constraints.
One can therefore solve the problem using a separation-based algorithm on the dual
(see Nisan and Segal(2001) for more details). 
This LP algorithm requires only a polynomial number of calls to a 
{\em demand oracle} that,
given a GS valuation $v$ and a price vector $\vec{p}$, computes a preferred set 
$S \in D(v \mid \vec{p})$ at these prices.

The fact that allocation among GS valuations can be computed 
in polynomial time follows also from the algorithms 
for calculating the convolution of $M^\#$-concave functions developed in
Murota(1996), Murota(2000) and Murota and Tamura(2001).
\begin{theorem}
\label{the:GSeasy}
An optimal allocation among GS valuations can be found in polynomial time, 
if the valuations themselves can
be computed in polynomial time.
\end{theorem}

A special case of interest is when all valuations 
in the auction are OXS valuations.  
If these valuations are given using the OR of XOR of
singletons representation, then the allocation problem reduces 
to a matching problem in a
bipartite graph and can thus be computed in polynomial time by 
Tarjan(1983), Fredman and Tarjan(1987).

\subsection{NP-hardness of an exact solution among submodular players}
\label{subsec:hard}
The problem of finding an optimal allocation in combinatorial auctions 
is known from Rothkopf and al.(1998) to be NP-hard 
even if bidders are single-minded, 
i.e., place only one bid each.
Is the problem any easier if the bidders are assumed to have submodular valuations?
The answer is negative. Finding an optimal allocation is still an NP-hard problem,
even if all valuations are additive with budget limit.
Note that, in such a case, all players valuations may be succinctly expressed.
\begin{theorem}
\label{the:hard}
Finding an optimal allocation in a combinatorial auction 
with two valuations that are
additive with a budget limit is NP-hard.
\end{theorem}

\begin{proof}
We will reduce from the well-known NP-complete 
problem ``Knapsack'' (see Garey and Johnson(1979)):
Given a sequence of integers $a_1 \ldots a_m$, 
and a desired total $t$, determine whether there
exists some subset $S$ of the integers whose sum is $t$, $\sum_{i \in S} a_i =t$.  
Given an input of this form, construct the following two valuations on $m$ items:  
\begin{itemize}
\item
The first valuation is additive giving the price $a_i$ to each item $i$: 
\mbox{$v_1(S) = \sum_{i \in S} a_i$}.
\item
The second valuation is additive with a budget limit of $2t$, and gives the price
$2a_i$ to each item $i$: \mbox{$v_2(S) = 2 \cdot \min(t,\sum_{i \in S} a_i)$}.
\end{itemize}

Fix an allocation of $S$ to valuation 2 and $S^c$ to valuation 1 
and consider the 3 cases: \mbox{$\sum_{i \in S} a_i < t$}, 
\mbox{$\sum_{i \in S} a_i = t$}, and 
\mbox{$\sum_{i \in S} a_i > t$}.  
Denote \mbox{$f = \sum_i a_i$}.
If \mbox{$\sum_{i \in S} a_i = t$} then \mbox{$t + \sum_{i \in S^c} a_i = f$} 
and
\mbox{$v_1(S^c) + v_2(S) =$} \mbox{$\sum_{i \in S^c} a_i + 2 t =$} \mbox{$f + t$}.
If \mbox{$\sum_{i \in S} a_i <$}  $t$ then
\mbox{$v_1(S^c) + v_2(S) =$} 
\mbox{$\sum_{i \in S^c} a_i + 2 \sum_{i \in S} a_i =$}
\mbox{$f + \sum_{i \in S} a_i <$} \mbox{$f + t$}.
If \mbox{$\sum_{i \in S} a_i >$} $t$ then \mbox{$\sum_{i \in S^c} a_i + t <$} $f$, and
\mbox{$v_1(S^c) + v_2(S) =$} \mbox{$\sum_{i \in S^c} a_i + 2t <$} \mbox{$f + t$}.  
Thus we see that the auction has an allocation $(S^c:S)$
with value $f+t$ if the knapsack problem has a positive answer $S$, 
and otherwise the allocation has a lower value. 
\end{proof}

Two comments should be made here: first, this NP-hardness result is 
for specific, succinctly defined, submodular valuations.  
The case of general submodular valuations given as an oracle is harder 
in a stricter sense, that of requiring exponentially many oracle queries 
as shown in Nisan and Segal(2001).  
Second, this result is for allocation among several valuations.  
A related well studied question is the maximization a single submodular function 
(not necessarily monotone). 
This problem is well known to be NP-hard for some specific succinctly defined 
functions, and to require exponentially many queries in the general oracle model,
see Jensen and Korte(1982), Lov\'{a}sz(1980), and Lov\'{a}sz(1983).

\subsection{A 2-approximation}
\label{subsec:2approx}
Our main algorithmic result is a $2$-approximation algorithm 
for combinatorial auctions in which
all valuations are submodular.
The result does not rely on any specific representation of submodular
functions: it only assumes that one can effectively compute valuations of subsets.  



\noindent
{\bf Input:} \mbox{$v_1, \ldots , v_n$} - submodular valuations,
given as black boxes.

\noindent
{\bf Output:} An allocation (partition of the items) \mbox{$S_1, \ldots , S_n$} 
which is a \mbox{$2$-approximation} 
to the optimal allocation

\noindent
{\bf Algorithm}
\begin{enumerate}
\item Set $S_1 = S_2 = \ldots = S_n \leftarrow \emptyset$.
\item For $x = 1 \ldots m$ do:

\begin{enumerate}
\item Let $j$ be the bidder with highest value of $v_j(x \mid S_j)$.
\item Allocate $x$ to bidder $j$, i.e. $S_j \leftarrow S_j \cup \{x\}$.
\end{enumerate}
\end{enumerate}
The algorithm obviously requires only a polynomial number of operations and
calls to valuation oracles for $v_j$ since
$v_j(x \mid S_j)=v_j(S_j \cup \{x\}) - v_j(S_j)$.

\begin{theorem}
\label{the:boundedapprox}
If all $v_i$ are submodular valuations then the
greedy algorithm above provides a \mbox{$2$}-approximation to the optimal one.
\end{theorem}
\begin{proof}
Let $Q$ be the original problem and define $Q'$ to be the problem on the $m-1$ 
remaining
items after item $1$ is removed: i.e., item 1 is unavailable and 
$v_j$ is replaced by ${v'}_j$ with 
${v'}_j(S) = v(S \mid \{1\}) = v(S \cup \{1\}) - v(\{1\})$,
where $j$ is the player to which item 1 was allocated.
All other valuations $v_{i}$, \mbox{$i \neq j$} are unchanged.
Notice that the algorithm above may be viewed as first allocating item $1$ to $j$ 
and then 
allocating the other elements using a recursive call on $Q'$.

Let us denote by $ALG(Q)$ the value of the allocation produced by this algorithm, 
and by $OPT(Q)$ the value of the optimal allocation. 
Let \mbox{$p = v_j(\{1\})$}.  
By the definition of $Q'$, it is clear that
\mbox{$ALG(Q)=ALG(Q') + p$}.  
We will now show that \mbox{$OPT(Q) \leq OPT(Q') + 2 p$}.
Let \mbox{$S_1, \ldots ,  S_n$} be the optimal allocation for $Q$,
and assume that \mbox{$1 \in S_k$}, i.e., item 1 is allocated to bidder $k$ 
by the algorithm above.
Let $S'$ be the solution to $Q'$ obtained from $S$ by ignoring item 1.
Let us compare the value of 
$S'$ in $Q'$(which is a lower bound on $OPT(Q')$) to the value of $S$ in $Q$ (which is exactly $OPT(Q)$).
All players except $k$ get the same allocation and all players except $j$ 
have the same valuation.  
Without loss of generality, assume \mbox{$k \neq j$}.
Player $k$ looses at most \mbox{$v_{k}(\{1\})$} since $v_k$ is submodular.
But \mbox{$v_{k}(\{1\}) \leq v_{j}(\{1\}) = p$} and
player $k$ looses at most \mbox{$p$}.
Player $j$ looses at most $p$ since, by monotonicity of $v_j$, 
$v'_{j}(S_j) = v_j(S_j \cup \{1\}) - v_j(\{1\}) \ge v_j(S_j) - p$.
Therefore \mbox{$OPT(Q') \geq$} \mbox{$OPT(Q) - 2 \: p$}.
The proof is concluded by induction on $Q'$ since, 
by lemma~\ref{lemma:smmargin}, $Q'$ also consists of 
submodular valuations:
\[
OPT(Q) \leq OPT(Q') + 2\: p \leq 
\]
\[
2 \: ALG(Q') + 2 \: p = 2 \: ALG(Q).
\]
\end{proof}
Looking at the proof one may see that many variants of the above algorithm 
also produce a $2$-approximation.
In particular the items $x$ may be enumerated in any order in the outer loop.
One may think of several heuristics for choosing the next $x$.
For example: take the item $x$ that maximizes the 
difference between the first and second values of $v_j(x \mid S_j)$.

An alternative proof for a variant of this algorithm 
may be obtained using the results
of Fisher and al.(1978) regarding the maximization of a single submodular 
valuation over a matroid.
Indeed the winner determination problem in a combinatorial auction is a
maximization problem over a matroid, and this does not seem to 
have been observed previously.
Consider the set \mbox{$P = N \times X$}, where $N$ is the set of agents and
$X$ is the set of items. 
A subset \mbox{$S \subseteq P$} is a partial allocation
iff each item is included in at most one element of $S$.
The family of partial allocations is a matroid.
The function to be maximized is the sum of 
\mbox{$n = \mid N \mid$} valuations.
In our case, each valuation is submodular and therefore their sum is also
submodular, and the existence of a polynomial-time $2$-approximation algorithm
follows from Theorem 2.1 of Fisher and al.(1978).
The NP-hardness result of section~\ref{subsec:hard} shows we are dealing with 
a hard case of submodular monotonic function maximization over a matroid.

The family of greedy algorithms we consider is wider than the one considered 
by those authors since they request a specific ordering of the items 
and we do not, leaving more space for heuristic considerations.
In section~\ref{sec:bounded} we shall generalize this result to valuations 
that are not submodular, and therefore fall outside 
the scope of Fisher and al.(1978).

Recent work (Lehmann(2002)) shows that the allocation
produced by this algorithm is not only a $2$-approximation to the optimal integer allocation but
also a $2$-approximation to the optimal {\em fractional} allocation.

It is easy to see that our bound is sharp and
the algorithm above may provide only a $2$-approximation 
even for submodular valuations:
take \mbox{$v_1(\{1\}) =$} \mbox{$v_1(\{2\}) =$} \mbox{$v_1(\{1 , 2 \}) =$}
\mbox{$v_2(\{1\}) =$} \mbox{$v_2(\{1 , 2\}) = 1$} and
\mbox{$v_2(\{2\}) = 0$}.

\vspace{0.1in}
\noindent
{\bf Open Problem:} 
Does any polynomial algorithm provide a better approximation
ratio?
\vspace{0.1in}

Conforti and Cornu\'{e}jols(1984) sharpened the results of Fisher and al.(1978).
Their results offer better bounds when the submodular valuations are known
to be close to linear, but not in the general case.
They also provide better bounds when the matroid is close to uniform, but 
our matroid has very small girth ($1$) and large rank ($m$).
We have also shown above that our bound is sharp.
The results of Nisan and Segal(2001)
provide a weak lower bound showing that achieving an approximation ratio of better than $1+1/m$ 
requires an exponential number of queries to the valuation oracles.

\subsection{Strategic considerations}
\label{sec:strategy}
The approximation algorithm presented above may be viewed 
as a simple sequence of auctions of the single items, one by one. 
Simplistic bidders whose strategies do not take the future
into account should indeed value an item $i$ at its current marginal value
for the bidder.  
Thus if each auction in this sequence is designed to be incentive compatible
(e.g. a second price auction) and if all bidders follow 
this simplistic bidding strategy (myopic bidders) then 
the desired $2$-approximate allocation would be obtained.  
In this subsection we ask whether there is a payment scheme 
to be used with this allocation algorithm 
that guarantees incentive-compatibility of the complete auction, 
i.e. that will reach this allocation when the bidders are rational
and not myopic.
The answer is negative.

An example will show that no payment scheme can make
the greedy allocation scheme 
for sub-modular combinatorial auctions a truthful mechanism.
In this example, the greedy scheme allocates the most expensive items first.
We do not know whether the result may be generalized to any greedy scheme.
The spirit of the proof is similar to that of Section 12
of Lehmann and al.(forthcoming).
\begin{example}
\label{ex:strategy}
Two goods: a and b. Two sub-modular players: Red and Green.
Red declares 10 for a, 6 for b and 11 for the set $\{a, b\}$.
Notice this is a sub-modular declaration, but this is not crucial.
Green has two personalities:
\begin{itemize}
\item Green1 declares 11 for a, 10 for b and 18 for the set $\{a, b\}$. 
\item Green2 declares 9 for a, 10 for b and 19 for the set $\{a, b\}$.
\end{itemize}
Notice both are sub-modular declarations.
\end{example}
The allocation between Red and Green1 goes in the following way:
a is allocated first to Green1 (Green's 11 vs. Red's 10) 
and then b is allocated to Green1 (Green's 18-11 vs. Red's 6).
Notice that if b had been allocated first Green would have obtained b but not a.
Green1 is therefore allocated the set $\{a, b\}$ and pays a sum $p$.
Notice that, since all declarations are fixed, $p$ is a number; it does not depend
on anything.
The allocation between Red and Green2 goes a different way: any ordering gives
a to Red and b to Green. Green2 pays $q$ (just a number).

If the mechanism is truthful and Green is Green1, it must be the case that Green
cannot gain by disguising himself as Green2: \mbox{$18 - p \geq 10 - q$}.
If the mechanism is truthful and Green is Green2, it must be the case that Green
cannot gain by disguising himself as Green1: \mbox{$10 - q \geq 19 - p$}.
A contradiction.

\section{False-Name Bids}
\label{sec:falsenames}
This section deals with
the following strategic manipulation problem 
that is inherent in combinatorial auctions 
that use the Vickrey-Clark-Groves payment rules 
(Vickrey(1961),Clarke(1971),Groves(1973) or see MacKie-Mason and Varian(1994)).
The VCG rules dictate that a player that is allocated a set $S$ of items 
pays the external cost to society of his allocation,
i.e., his payment is $u(X) - u(X-S)$, where $X$ is the set of all items, 
and $u$ is the combined (by OR) valuations of all other
bidders (i.e. the valuation obtained by the value of the optimal allocation 
of items among the other bidders).
A disturbing observation made in Sakurai and al.(1999) is that 
in many cases a bidder can manipulate a VCG combinatorial
auction and reduce his payment by splitting his bid and placing 
two separate bids under {\em false-names}.
For example, if two items $A$ and $B$ are offered and my valuation 
for the pair $\{A,B\}$ is $6$, 
while another bidder values the pair at $5$, the VCG rules will set 
my payment to $5$.  
If, instead, I place two separate bids, $\{A\}$ for $4$ and
$\{B\}$ for $4$, then one may easily calculate that each 
of these {\em false-name bidders} pays $1=5-4$ for a total
payment of $2$. 

This issue was further discussed in Yokoo and al.(2000a, 2000b, 20001). 
They showed that such false-name bidding cannot be profitable for any agent 
if the surplus function is concave, i.e., submodular over agents. 
This condition is a strengthening of the
{\em agents are substitutes} condition first considered in Shapley(1962)
and then in Bikhchandani and Ostroy(forthcoming).
We show that a different condition also
guarantees that false-name bidding is not profitable. 
If the aggregate valuation of all other agents is submodular
{\em in the items}, one cannot benefit from placing false-name bids.
We then discuss this condition and show that the result is sharp:
false-name attacks may be profitable even if all agents are submodular.
\begin{theorem}
\label{the:false}
A player cannot benefit from placing false name bids 
in a combinatorial auction using the VCG rules,
whenever the combined valuation of all other players is submodular.
\end{theorem}
\begin{proof}
We shall show that two players, one of them bidding a valuation
$v_1$ and the other a valuation $v_2$ will pay {\em no less} than a
single player bidding \mbox{$v_1 \vee v_2$}, as long as the combined
valuation function of the other players is submodular.

Consider the set of bids \mbox{$v_1, v_2, u_1, \ldots , u_m$} and let $S_1$ be
the set allocated to player 1, $S_2$ be the set allocated to
player 2, and $T$ be the set allocated to all the $u_i$ bidders
together: thus the total set of goods is \mbox{$T \cup S_1 \cup S_2$}. 
Let \mbox{$u = u_1 \vee \ldots \vee u_m$}. 

The VCG rules (GVA auction) specify that player 1 will pay: 
$$(u \vee v_2)(T \cup S_1 \cup S_2) - (u \vee v_2)(T \cup S_2).$$ 
Since the optimal allocation of 
$T \cup S_2$ among the valuations $u$ and $v_2$ allocates $T$ to $u$ and
$S_2$ to $v_2$, we have that 
\mbox{$(u \vee v_2)(T \cup S_2) =$} \mbox{$u(T)+v_2 (S_2 )$}.  
By considering the allocation of $S_2$ to $v_2$
and $T \cup S_1$ to $u$, we can bound 
\[
(u \vee v_2)(T \cup S_1 \cup S_2) \ge u(T \cup S_1) + v_2(T_2).
\]  
We thus get that player 1 pays at least \mbox{$u(T \cup S_1) - u(T)$}. 
Similarly player 2 will pay at least 
\mbox{$u(T \cup S_2) - u(T)$}.

Consider, on the other hand what happens when instead of $v_1$ and
$v_2$ a single \mbox{$v_1 \vee v_2$} bid is submitted.  
The allocation to this bidder is exactly \mbox{$S_1 + S_2$}.  
The VCG payment of this player will be \mbox{$u(T \cup S_1 \cup S_2) - u(T)$}. 
The submodularity of $u$ directly implies that this is less than or equal to the sum of
the payments of players 1 and 2 in the previous case.
\end{proof}

As a simple corollary we obtain.
\begin{corollary}
\label{co:false}
A player cannot benefit from placing false name bids 
in a combinatorial auction using the VCG rules,
where all valuations satisfy the gross substitutes property.
\end{corollary}

\begin{proof}
According to theorem \ref{OGS}, 
the combined valuation (OR) of GS valuations is GS and thus is also submodular.
\end{proof}
A slightly weaker result could have been proven by showing that the aggregate
utility (OR combination) of GS valuations is concave in the agents and using 
Proposition 3 of Yokoo and al.(2000a).
But Common Knowledge of the fact that all players are GS would have been required.

Corollary~\ref{co:false} also implies a previous 
result of Sakurai and al.(1999) 
showing that in a Vickrey multi-unit auction 
of identical items in which all players have a downward sloping valuations, 
no player can benefit from placing false-name bids. 

Could Theorem~\ref{the:false} assume, instead, that all players have a submodular
valuation? The following example shows that this is not the case.
The following example builds on Example~\ref{ex:OR} and presents an auction in which
each player is submodular, but the combined valuation of the opponents is not and
one can benefit from false-name bids.  

\begin{example}
\label{ex:false}
Red has budget limit 6 and values: \mbox{$red(a) =$} \mbox{$red(c) = 3$},
\mbox{$red(b) = 5$}.
Blue has unbounded budget: \mbox{$blue(a) = 1$}, \mbox{$blue(b) = 2$},
\mbox{$blue(c) = 0$}.
Let $v$ be the combined valuation, \mbox{$v = red \vee blue$}.
We have that \mbox{$v(abc) = 8$}, \mbox{$v(ab) =$} 
\mbox{$v(ac) =$} \mbox{$v(bc)=6$}.
Green has unbounded budget: \mbox{$green(a) = 2$}, $green(b) = 0$,
\mbox{$green(c) = 5$}.
\end{example}
If Green acts as himself, 
he gets $\{a, c\}$ (utility 7) and Red gets $\{b\}$ (utility 5) for a
total of 12, unbeatable.  
Green pays: 8-5 = 3.

But, assume that Green acts under two different identities 
$G1(a) = 2$ and $G2(c)=5$.
Obviously $G1$ gets $\{a\}$ and $G2$ gets $\{c\}$. 
$G1$ pays: 11-10=1. $G2$ pays: 8-7=1. On the whole $G1$ and $G2$ together pay
only 2 which is strictly less than 3.

This example also shows that the 
``agents are substitutes'' condition of Shapley(1962) and Bikhchandani and 
Ostroy(forthcoming) 
does {\em not} follow from the fact that all agents have submodular valuations.
Let $T$ be the set of agents containing: Red, Blue, G1 and G2.
We have seen that \mbox{$v(T) = 12$}. Let $S$ be the set containing G1 and G2.
We have seen that \mbox{$v(T-S) = 8$}.
But \mbox{$v(T-\{G1\}) = 11$} and \mbox{$v(T-\{G2\}) = 8$}.
We see that \mbox{$v(T)-v(T-S) < (v(T)-v(T-\{G1\})) + (v(T)-v(T-\{G2\}))$}.

\section{Bounded Complementarity}
\label{sec:bounded}
While the definition of submodular valuations is theoretically clean, in many
{\em real world} situations, valuations cannot be assumed to be submodular.
In many cases, valuations exhibit complementarities. Complementarities are,
indeed, one of the main reasons for the development of parallel or combinatorial
auctions. 
In much of traditional economic theory, decreasing marginal values,
or convexity, have been assumed more for mathematical expediency reasons than 
because they fit the data.
The submodularity assumption will now be weakened in a way that allows it to
encompass complementarities and nevertheless retains its mathematical structure.

In many cases it can be argued that marginal valuations should be 
{\em rather} decreasing, i.e., should not be increasing {\em too much},
or that if complementarities exist, they are not too large and may be bounded 
in some way.  
In this section we formalize this idea and then observe that 
in such cases the greedy allocation algorithm still produces provably good results.
Thus our results assuming submodularity have wider implications. 
 
The definitions and results of Section~\ref{sec:submodular} 
can be generalized to deal
with valuations that exhibit a limited amount of complementarity.
We assume $a \geq 1$ is a real number.
The number $a$ is a measure of how close we are to complement-freeness or to
submodularity.
The case $a = 1$ is the one considered above in this paper: complement-freeness
and submodularity. The larger the number $a$ the looser our assumptions.
\begin{definition}
\label{anocompl-def}
A valuation $v$ is said to exhibit $a$-bounded complementarities if, 
for any set $A$
and item $x$: 
\[
v(A \cup \{ x \}) \leq v(A) + a \: v(\{ x \}).
\]
\end{definition}
Complementarities are bounded multiplicatively by the number $a$.
\begin{definition}
\label{abound-def}
A valuation $v$ is said to be $a$-submodular
if and only if for every subset $W$ of items, 
the marginal valuation $v(\cdot \mid W)$ exhibits $a$-bounded complementarities.
\end{definition}
The proofs of the following results are similar to those of 
Section~\ref{sec:submodular} and will be omitted.
\begin{proposition}
\label{asm-def}
A valuation $v$ is $a$-submodular if and only if 
any one of the following 
equivalent propositions holds.
\begin{itemize}
\item 
For any \mbox{$x \in X$} and \mbox{$S, T \subseteq X$}, such that
\mbox{$S \subseteq T$} and \mbox{$x \not \in T$}: 
\mbox{$a \: v(x \mid S) \geq v(x \mid T)$}.
\item For any \mbox{$S, T, V \subseteq X$}, such that
\mbox{$S \subseteq T$}: \mbox{$a \: v(V \mid S) \geq v(V \mid T)$}.
\item For any \mbox{$A, B \subseteq X$}:
\mbox{$v(A) + a \: v(B) \geq v(A \cup B) + a \: v(A \cap B)$}.
\end{itemize}
\end{proposition}

The $2$-approximation result of Section~\ref{sec:allocation} generalizes 
to a $1+a$-approximation.
\begin{theorem}
\label{the:a-approx}
The greedy algorithm provides a $(1+a)$ approximation to the optimal one
if all valuations are $a$-submodular.
\end{theorem}
Notice that the problem we solved is of maximizing a function that is 
{\em not} submodular over a matroid.
We do not know whether this may have useful implications for other such
problems.

\section{Acknowledgments}
We thank Sushil Bikhchandani, Motti Perry, Rakesh Vohra and Makoto Yokoo for helpful discussions,
and Liad Blumrosen for comments on an earlier draft of this manuscript.
An associate editor pointed to us the relevant work on optimization of submodular
functions on matroids.
\bibliographystyle{plain}

\begin{thebibliography}{10}

\bibitem{AusubelDynamic:unp}
Ausubel, L.~M.(2000).
\newblock ``An Efficient Dynamic Auction for Heterogeneous Commodities,''
\newblock Working paper.

\bibitem{BVSV:unp}
Bikhchandani, S., de~Vries, S., Schummer, J., and Vohra, R.~V.(2002).
\newblock ``Linear Programming and {Vickrey} Auctions,''
\newblock in {\em Mathematics of the Internet: E-auction and Markets} 
(B.~Dietrich, R.~V.~Vohra, and P.~Brick, Eds.), 
IMA Series in Mathematics and its Applications. Springer.

\bibitem{BikhOstroy:JET}
Bikhchandani, S., and Ostroy, J.~M.(forthcoming).
\newblock ``The Package Assignment Model,''
\newblock {\em Journal of Economic Theory}.

\bibitem{Clarke:71}
Clarke, E.~H.(1971).
\newblock ``Multipart Pricing of Public Goods,''
\newblock {\em Public Choice}, 11:17--33.

\bibitem{ConfCorn:84}
Conforti, M., and Cornu\'{e}jols, G.(1984).
\newblock "Submodular Set Functions, Matroids
and the Greedy Algorithm: Tight Worst-Case Bounds and Some
Generalizations of the Rado-Edmonds Theorem,"
\newblock {\em Discrete Applied Mathematics}, 7:251-274.

\bibitem{deVriesVohra:survey}
de~Vries, S., and Vohra, R.~V.(forthcoming).
\newblock ``Combinatorial Auctions: A Survey,''
\newblock {\em INFORMS Journal of Computing}.
\newblock Available from http://www.kellogg.nwu.edu/faculty/vohra/htm/res.htm,
  Test problems available from:
  http://www-m9.mathematik.tu-muenchen.de/\~devries/comb\_auction\_supplement/.

\bibitem{FishNemWol:78}
Fisher, M.~L., Nemhauser, G.~L., and Wolsey, L.~A.(1978).
\newblock ``An Analysis of Approximations for Maximizing Submodular Set Functions-II,''
\newblock in {\em Polyhedral Combinatorics}, pp. 73--87 Mathematical Programming Studies 8. North-Holland, Amsterdam.

\bibitem{FredTar:87}
Fredman, M.~L., and Tarjan, R.~E.(1987).
\newblock ``Fibonacci Heaps and their Uses in Improved Network Optimization
  algorithms,''
\newblock {\em Journal of the ACM}, 34(3):596--615.

\bibitem{FujiYang:01}
Fujishige, S., and Yang, Z.(2001).
\newblock ``A Note on the Relationship between Kelso and Crawford's Gross
  Substitutes Condition and M\#-concave Functions,''
\newblock Preprint.

\bibitem{FLBS:99}
Fujishima, Y., Leyton-Brown, K., and Shoham, Y.(1999).
\newblock ``Taming the Computational Complexity of Combinatorial Auctions:
  Optimal and Approximate Approaches,''
\newblock in {\em Proceedings of IJCAI'99}, Stockholm, Sweden.
  Morgan Kaufmann.

\bibitem{GareyJohnson:79}
Garey, M.~R., and Johnson, D.~S.(1979).
\newblock {\em Computers and Intractability: a Guide to the Theory of
  NP-completeness}.
\newblock Freeman, San Francisco.

\bibitem{GonLeh:EC00}
Gonen, R., and Lehmann, D.(2000).
\newblock ``Optimal Solutions for Multi-unit Combinatorial Auctions: Branch and
  Bound Heuristics,''
\newblock in {\em Second ACM Conference on Electronic Commerce (EC-00)}, pp.
  13--20, Minneapolis, Minnesota.

\bibitem{Groves:73}
Groves, T.(1973).
\newblock ``Incentives in Teams,''
\newblock {\em Econometrica}, 41:617--631.

\bibitem{GulStacc:99}
Gul, F., and Stacchetti, E.(1999).
\newblock ``Walrasian Equilibrium with Gross Substitutes,''
\newblock {\em Journal of Economic Theory}, 87:95--124.

\bibitem{GulStacc:00}
Gul, F. and Stacchetti, E.(2000).
\newblock ``The English Auction with Differentiated Commodities,''
\newblock {\em Journal of Economic Theory}, 92:66--95.

\bibitem{JensenKorte:82}
Jensen, P.~M., and Korte, B.(1982).
\newblock ``Complexity of Matroid Property Algorithms,''
\newblock {\em SIAM Journal on Computing}, 11:184--190.

\bibitem{KelsoCraw:82}
Kelso, A.~S., and Crawford, V.~P.(1982).
\newblock ``Job Matching, Coalition Formation and Gross Substitutes,''
\newblock {\em Econometrica}, 50:1483--1504.

\bibitem{LCS:EC99}
Lehmann, D., O'Callaghan, L.~I., and Shoham, Y.(forthcoming).
\newblock ``Truth Revelation in Approximately Efficient Combinatorial
  Auctions,''
\newblock {\em Journal of the ACM}.
\newblock A preliminary version appeared in 
{\em Proceedings of the First ACM Conference on Electronic
  Commerce. EC'99}, pp. 96--102, Denver, Colorado. SIGecom,
  ACM Press.

\bibitem{Lehmann:Dagstuhl}
Lehmann, D.(2002).
\newblock ``What is there when there is no Walrasian Equilibrium?''
\newblock {\em presented at workshop on Electronic Market Design}. Dagstuhl, 
Germany.

\bibitem{Lovasz:80}
Lov\'{a}sz, L.(1980).
\newblock ``Matroid Matching and some Applications,''
\newblock {\em Journal of Combinatorial Theory (B)}, 28:208--236.

\bibitem{Lovasz:83}
Lov\'{a}sz, L.(1983).
\newblock ``Submodular Functions and Convexity,''
\newblock in {\em Mathematical Programming -- The State of the Art}
(A.~Bachem, M.~Gr\"{o}tschel, and B.~Korte, Eds), pp. 235--257. Springer-Verlag.

\bibitem{VarianMacK:GVA}
MacKie-Mason, J.~K., and Varian, H.~R.(1994).
\newblock ``Generalized {Vickrey} Auctions,''
\newblock Working paper, Un. of Michigan, available at 
http://www-personal.umich.edu/~jmm.

\bibitem{Milgrom:2000}
Milgrom, P.(2000).
\newblock ``Putting Auction Theory to Work: the Simultaneous Ascending Auction,''
\newblock {\em Journal of Political Economy}, 108(2):245--272.

\bibitem{Moulin:book88}
Moulin, H.(1988).
\newblock {\em Axioms of Cooperative Decision Making}.
\newblock Cambridge University Press, Cambridge, U.K.

\bibitem{Murota:96}
Murota, K.(1996).
\newblock ``Valuated Matroid Intersection, II: Algorithm,''
\newblock {\em SIAM Journal on Discrete Mathematics}, 9:562--576.

\bibitem{Murota:Matrices}
Murota, K.(2000).
\newblock {\em Matrices and Matroids for Systems Analysis}.
\newblock Springer-Verlag.

\bibitem{MurotaShioura:99}
Murota, K., and Shioura, A.(1999).
\newblock ``M-convex Functions on Generalized Polymatroid,''
\newblock {\em Mathematics of Operations Research}, 24:95--105.

\bibitem{MurotaTamura:01}
Murota, K. and Tamura, A.(2001).
\newblock ``Applications of M-convex Submodular Flow Problem to Mathematical 
Economics,''
\newblock in {\em Algorithms and Computation, ISAAC2001}
(P.~Eades, and T.~Takaoka, Eds.), pp 14--25.
Lecture Notes in Computer Science 2223, Springer.

\bibitem{Narayanan:97}
Narayanan, H.(1997).
\newblock {\em Submodular Functions and Electrical Networks}.
\newblock Volume~54 of Annals of Discrete Mathematics, North-Holland.

\bibitem{Nisan:EC00}
Nisan, N.(2000).
\newblock ``Bidding and Allocation in Combinatorial Auctions,''
\newblock in {\em Proceedings of the 2nd ACM Conference on Electronic Commerce
  EC'00}, pp. 1--12, Minneapolis, Minnesota. ACM Press.

\bibitem{NisanSegal:01}
Nisan, N., and Segal, I.(2001).
\newblock ``The Communication Complexity of Efficient Allocation Problems,''
\newblock available at http://www.cs.huji.ac.il/~noam.

\bibitem{RothPekHar}
Rothkopf, M.~H., Peke\v{c}, A., and Harstad, R.~M.(1998).
\newblock ``Computationally Manageable Combinatorial Auctions,''
\newblock {\em Management Science}, 44(8):1131--1147.

\bibitem{YokooAAAI:99}
Sakurai, Y., Yokoo, M., and Matsubara, S.(1999).
\newblock ``A Limitation of the Generalized Vickrey Auction in Electronic
  Commerce : Robustness Against False-name Bids,''
\newblock in {\em Sixteenth National Conference on Artificial Intelligence
  (AAAI-99)}, pp. 86--92.

\bibitem{Sandholm:AAAI99}
Sandholm, T., and Suri, S.(1999).
\newblock ``Improved Algorithms for Optimal Winner Determination in Combinatorial
  Auctions and Generalizations,''
\newblock in {\em Proceedings of the National Conference on Artificial
  Intelligence (AAAI)}, Austin, TX.

\bibitem{Sandholm99}
Sandholm, T.(1999).
\newblock ``An Algorithm for Optimal Winner Determination in Combinatorial
  Auctions,''
\newblock in {\em IJCAI-99}, pp. 542--547, Stockholm.

\bibitem{Shapley:subst}
Shapley, L.(1962).
\newblock ``Complements and Substitutes in the Optimal Assignment Problem,''
\newblock {\em Naval Research Logistics Quarterly}, 9:45--48.

\bibitem{Tarj:Data}
Tarjan, R.~E.(1983).
\newblock {\em Data Structures and Network Algorithms}.
\newblock Society for Industrial and Applied Mathematics, Philadelphia, PA.

\bibitem{Tenn:sometractauc}
Tennenholtz, M.(2002).
\newblock ``Tractable Combinatorial Auctions and B-matching,''
\newblock {\em Artificial Intelligence}, 140:231--243.
\newblock Preliminary version in {\em Seventeenth National Conference on 
Artificial Intelligence (AAAI-2000), pp. 98--103}.

\bibitem{Topkis:supermod}
D.~M.~Topkis(1998).
\newblock {\em Supermodularity and Complementarity}.
\newblock Princeton University Press, 1998.

\bibitem{Varian:95}
Varian, H.~R.(1995).
\newblock ``Economic Mechanism Design for Computerized Agents,''
\newblock in {\em Proceedings of the First Usenix Conference on Electronic
  Commerce}, New York.

\bibitem{Vickrey:61}
Vickrey, W.~S.(1961).
\newblock ``Counterspeculation, Auctions and Competitive Sealed Tenders,''
\newblock {\em Journal of Finance}, 16:8--37.

\bibitem{YokooICDS:00}
Yokoo, M., Sakurai, Y., and Matsubara, S.(2000a).
\newblock ``The Effect of False-name Declarations in Mechanism Design: Towards
  Collective Decision Making on the Internet,''
\newblock in {\em 20th International Conference on Distributed Computing
  Systems (ICDCS-2000)}, pp. 146--153.

\bibitem{YokooAAAI:00}
Yokoo, M., Sakurai, Y., and Matsubara, S,(2000b).
\newblock ``Robust Combinatorial Auction Protocol Against False-Name Bids,''
\newblock in {\em Seventeenth National Conference on Artificial Intelligence
  (AAAI-2000)}, pp. 110--116.

\bibitem{YokooICDCS:01}
Yokoo, M., Sakurai, Y., and Matsubara, S.(2001).
\newblock ``Robust Double Auction Protocol Against False-name Bids,''
\newblock in {\em 21st International Conference on Distributed Computing
  Systems (ICDCS-2001)}.

\end{thebibliography}

\end{document}